\RequirePackage{lineno}
\documentclass[prd,twocolumn,amsmath,amssymb,showpacs,superscriptaddress,nofootinbib]{revtex4-1}

\usepackage{graphicx}  
\usepackage{dcolumn}   
\usepackage{bm}        
\usepackage{amssymb}   
\usepackage{amsfonts}  
\usepackage{hyperref}
\usepackage{amsmath}
\usepackage{mathtools}
\usepackage{mathrsfs}

\begin{document}

\title{\bf Study of the Dalitz decay \boldmath{$J/\psi \rightarrow e^+e^- \eta$} }
\author{\small
M.~Ablikim$^{1}$, M.~N.~Achasov$^{9,d}$, S. ~Ahmed$^{14}$, M.~Albrecht$^{4}$, M.~Alekseev$^{55A,55C}$, A.~Amoroso$^{55A,55C}$, F.~F.~An$^{1}$, Q.~An$^{52,42}$, Y.~Bai$^{41}$, O.~Bakina$^{26}$
% ...
\\
\vspace{0.2cm}
(BESIII Collaboration)\\
\vspace{0.2cm} {\it
$^{1}$ Institute of High Energy Physics, Beijing 100049, People's Republic of China\\
% ...
$^{4}$ Bochum Ruhr-University, D-44780 Bochum, Germany\\
% ...
$^{9}$ G.I. Budker Institute of Nuclear Physics SB RAS (BINP), Novosibirsk 630090, Russia\\
% ...
$^{14}$ Helmholtz Institute Mainz, Johann-Joachim-Becher-Weg 45, D-55099 Mainz, Germany\\
% ...
$^{26}$ Joint Institute for Nuclear Research, 141980 Dubna, Moscow region, Russia\\
% ...
$^{41}$ Southeast University, Nanjing 211100, People's Republic of China\\
$^{42}$ State Key Laboratory of Particle Detection and Electronics, Beijing 100049, Hefei 230026, People's Republic of China\\
% ...
$^{52}$ University of Science and Technology of China, Hefei 230026, People's Republic of China\\
% ...
$^{55}$ (A)University of Turin, I-10125, Turin, Italy; (B)University of Eastern Piedmont, I-15121, Alessandria, Italy; (C)INFN, I-10125, Turin, Italy\\
% ...
$^{d}$ Also at the Novosibirsk State University, Novosibirsk, 630090, Russia\\
% ...
}\vspace{0.4cm}}

\begin{abstract}
  We study the electromagnetic Dalitz decay $J/\psi \to e^+e^- \eta$ and search for di-electron decays of a dark gauge boson ($\gamma'$) in  $J/\psi \to \gamma' \eta$  with the two $\eta$ decay modes $\eta \rightarrow \gamma \gamma$ and $\eta \rightarrow \pi^+\pi^-\pi^0$ using  $(1310.6\pm 7.0)\times10^6$  $J/\psi$ events collected with the BESIII detector. 
The branching fraction of $J/\psi \to e^+e^- \eta$ is measured to be $(1.42 \pm 0.04 ({\rm stat}) \pm 0.07 ({\rm syst}))\times 10^{-5}$, with a precision that is improved by a factor of $1.5$ over the previous BESIII measurement. The corresponding di-electron invariant mass dependent modulus square of the transition form factor is explored for the first time, and the pole mass is determined to be $\Lambda = 2.56 \pm 0.04({\rm stat}) \pm 0.03({\rm syst})$ GeV/$c^2$.  We find no evidence of $\gamma'$ production and set $90\%$ confidence level upper limits on the product branching fraction  
$\mathcal{B}(J/\psi \to \gamma' \eta)\times \mathcal{B}(\gamma' \to e^+e^-)$ as well as the kinetic mixing strength  between the Standard Model photon and $\gamma'$ in the mass range of  $0.01 \le m_{\gamma'} \le 2.4$ GeV/$c^2$.

\end{abstract}


\maketitle

\section{Introduction}
The study of electromagnetic (EM) Dalitz decays of a vector meson ($V=\rho,\omega,\phi,J/\psi$) into a pseudoscalar meson ($P=\pi^0,\eta,\eta'$) and a lepton-pair, $V \rightarrow \ell^+\ell^- P$ ($\ell = e,\mu$), plays an important role in revealing the structure of hadrons and the interaction mechanism between photons and hadrons~\cite{Landsberg}. These decays proceed via $V \rightarrow \gamma^*P$ in which  the virtual photon $\gamma^*$ subsequently converts  into a lepton pair. Assuming the mesons to be pointlike particles,  the di-lepton invariant mass ($m_{\ell^+\ell^-}$)  dependent decay rate of $ V \rightarrow \ell^+\ell^- P$ can be described by 
quantum electrodynamics (QED)~\cite{Kroll}. Any deviation from the QED prediction, caused by the dynamics of the EM structure arising at the $V \rightarrow P$ transition vertex, is formally described by a transition form factor (TFF)~\cite{Landsberg}.  The dependence of the differential decay rate of $V \rightarrow P \ell^+\ell^-$  on the four-momentum transfer squared $q^2 = m_{\ell^+\ell^-}^2$ is parameterized as~\cite{Landsberg}
\begin{widetext}
\begin{eqnarray}
  \frac{d\Gamma(V \rightarrow P \ell^+\ell^-)}{dq  \Gamma (V \rightarrow P \gamma)}
   &=&
   \frac{2\alpha}{3\pi   q} \left(1-\frac{4m^2_\ell}{q^2}\right)^{1/2} \left(1+\frac{2m^2_\ell}{q^2}\right) \left[\left(1+\frac{q^2}{m^2_{V}-m^2_P}\right)^2 - \frac{4m^2_{V}q^2}{(m^2_{V}-m^2_P)^2}\right]^{3/2} \times |F_{VP}(q^2)|^2
 \nonumber \\
 & =& [\mbox{QED}(q^2)]  \times |F_{VP}(q^2)|^2,
 \label{eq:dgamman}
\end{eqnarray}
\end{widetext}

\noindent where $\mbox{QED}(q^2)$ is the QED predicted $q^2$ dependent decay rate,   $\alpha$  is the fine structure constant, $F_{VP}(q^2)$ is the $q^2$ dependent TFF, and $m_\ell$, $m_{V}$ and $m_{P}$ are the masses of leptons, $V$ and $P$ mesons, respectively.

The TFFs of light mesons contribute to the hadronic light-by-light (HLBL) corrections~\cite{Colangelo} to the theoretical determination of the muon anomalous magnetic moment, $a_{\mu}= (g_{\mu}-2)/2$, which provides a low-energy test of the completeness of the Standard Model (SM)~\cite{Jegerlehner,Nyffeler}. Experimentally, it can directly be accessible by comparing the $m_{\ell^+\ell^-}$ spectrum of Dalitz decays $V \rightarrow \ell^+\ell^- P$  with that of the point-like QED prediction~\cite{Landsberg}. Within the vector meson dominance model (VMD)~\cite{Sakurai}, the TFF is mainly governed by the coupling of the $\gamma^*$ to the $V$ meson via an intermediate vector ($V'$) meson in the timelike region, and is commonly expressed as a multipole function in the charmonium mass region~\cite{Budnev}
\begin{equation}
F_{VP}(q^2) = N\sum_{V'} A_{V'} \frac{m_{V'}^2}{m_{V'}^2-q^2-i\Gamma_{V'} m_{V'}},
\label{TFFcurve}
\end{equation}

\noindent where $N$ is a normalization constant ensuring that $F_{VP}(0)=1$, $V'$ denotes the intermediate resonances $\rho,\omega$, $\phi$, and charmonium vector mesons, $m_{V'}$, $\Gamma_{V'}$ and $A_{V'}$ are the corresponding masses, widths and the coupling constants. The contribution of vector mesons with masses above ($m_V-m_P$), {\it non-resonant} contribution, is often represented as
\begin{equation}
F_{VP} (q^2) = \frac{1}{1-q^2/\Lambda^2},
\label{TFF}
\end{equation}

\noindent where $\Lambda$ is an effective pole mass. The inverse square value  $\Lambda^{-2}$ reflects the slope of the TFF at $q^2 = 0$.

 The EM Dalitz decays of  light unflavored vector mesons $\rho$, $\omega$ and $\phi$ are well established by several collider and nuclear physics experiments~\cite{CMD2, NA60, SND, KLOE2,mami}. The BESIII collaboration reported the first measurements of the branching fractions of $J/\psi \to e^+e^- P$ and the TFF of $J/\psi \to e^+e^- \eta'$ using a data sample of 225 million $J/\psi$ events~\cite{xinkun}. The results agree well with the VMD predictions based on a simple pole approximation~\cite{haibo} within the statistical uncertainties. 
BESIII has recently accumulated $5$ times more statistics of the $J/\psi$ data-set~\cite{njps}, which can be used to improve the precision of these measurements and enable measurement of the TFFs of $J/\psi \rightarrow e^+e^- P$.

 The EM Dalitz decays can also be utilized to search for a hypothetical dark photon, $\gamma'$, via the decay chain $J/\psi \to \gamma' P$, $\gamma' \rightarrow \ell^+\ell^-$~\cite{haibo,Reece}. The $\gamma'$ is a new type of force carrier in the simplest scenario of an Abelian $U(1)$ interaction under which dark matter particles are considered to be charged~\cite{Arkani,Finkbeiner,Pospelov}. A $\gamma'$ with mass below twice the proton mass can explain the features of the electron/positron excess observed 
  by the cosmic ray experiments~\cite{integral,Adriani,Ackermann,Aguilar}. A dark photon with such a low mass can also explain the presently observed deviation of $a_{\mu}$ up to the level of $(3 -4 )\sigma$ between the measurement and SM prediction~\cite{Pospelov}. The $\gamma'$ couples with the SM photon through its kinetic mixing with the SM hypercharge field~\cite{Holdom}. The coupling strength between the dark sector and the SM, $\epsilon$, is parameterized as $\epsilon^2 = \alpha'/\alpha$, where $\alpha'$ is the fine structure constant in the dark sector. A series of experiments have reported null results in $\gamma'$ searches, including the $a_{\mu}$ favored region, and have constrained the $\epsilon$ values as a function of $\gamma'$ mass to be below $10^{-3}$~\cite{babar,NA48,bes}. More experimental information about the $\gamma'$ searches via new decay modes, such as $J/\psi \to \gamma' P$, might be helpful to understand some other possible scenarios of the $\gamma'$ coupling to the SM~\cite{fayet}. In this paper, we present a study of the EM Dalitz decays $J/\psi \to e^+e^- \eta$ and search for di-electron decays of a dark photon through $J/\psi \to \gamma'\eta$ using $(1310.6 \pm 7.0)\times 10^6$ $J/\psi$ events collected with the BESIII detector~\cite{njps}.

 \section{The BESIII experiment and Monte Carlo simulation}
 \label{MCsimulation}
The BESIII detector is a general purpose spectrometer containing four major detector sub-components with a geometric acceptance of $93\%$ of the total solid angle as described in Ref.~\cite{bes3detector}. A helium-based ($60\%$ He, $40\%$ $C_3H_8$) multi-layer drift chamber (MDC), which contains 43 layers and operates in a 1.0~T (0.9~T) solenoidal magnetic field for the 2009 (2012) $J/\psi$ data, is used to measure the momentum of the charged particles. Charged particle identification (PID) is based on the energy loss (d$E$/d$x$) in the tracking system and the time-of-flight (TOF) measured by a scintillation based TOF detector containing one barrel and two end-caps. A CsI(Tl) based electromagnetic calorimeter (EMC) is used to measure the energies of photons and electrons, while a muon counter containing nine (eight) layers of resistive plate chamber counters interleaved with steel in the barrel (end-cap) region is used for muon identification.

Monte Carlo (MC) simulated events are used  to optimize the event selection criteria, to study the detection acceptance and to understand the potential backgrounds. The {\sc Geant4}~\cite{geant4} based simulation package contains the information about the detector geometry and material description, the detector response and signal digitization models, as well as the records of time dependent detector running conditions and performance. An  MC sample of $1.225$ billion inclusive $J/\psi$ decays is generated for the background studies with the {\sc EvtGen} generator~\cite{evtgen} for the known $J/\psi$ decay modes with the branching fractions set to their world average value taken from Ref.~\cite{pdg}, and the {\sc lundcharm} package~\cite{lundcharm} for the remaining unknown $J/\psi$ decay modes. The {\sc kkmc} event generator package~\cite{kkmc} is used to simulate the production of the $J/\psi$ resonance via $e^+e^-$ annihilation, incorporating the effects of the beam energy spread and initial-state-radiation (ISR).

The angular distribution of the decay $J/\psi \to e^+e^- \eta$ is simulated according to a combined formula of Eqs.~(4) and (6) of Ref.~\cite{haibo}, where the dependence on the cosine of the $\eta$ meson polar angle in the $J/\psi$ rest frame ($\cos\theta_{\eta}$) is parameterized by $(1+\alpha_{\theta} \cos^2\theta_{\eta})$ with $\alpha_{\theta}=1.0$ measured from the data as described in Section~\ref{tffmeasurement} to take into account of the $J/\psi$ polarization state in the $e^+e^-$ annihilation system, and the TFF is assumed to follow Eq.~(\ref{TFF}) with  $\Lambda = 2.56$~GeV/$c^2$ measured in this analysis also described in Sec.~\ref{tffmeasurement}. The $J/\psi \rightarrow \gamma' \eta$ decay is modeled by a helicity amplitude model and $\gamma' \rightarrow e^+e^-$ decay by a model of a vector meson decaying to a lepton-pair~\cite{evtgen}.

\section{Data analysis}
\label{Data_sample}
In this analysis, the $\eta$ meson candidates are reconstructed using the dominant decay modes $\eta\to\gamma \gamma$ and $\eta\to\pi^+\pi^-\pi^0$,
where the $\pi^0$ meson is reconstructed with a $\gamma \gamma$ pair. We select events of interest with two (four) charged tracks with zero net charge in the $\eta \rightarrow \gamma \gamma$ ($\eta \rightarrow \pi^+\pi^-\pi^0$) decay and at least two good photon candidates. The charged tracks are required to be measured in the active region of the MDC, $|\cos\theta| < 0.93$, where $\theta$ is the polar angle of the charged tracks. They must also have the points of closest approach to the beam line within $\pm 10.0$~cm from the interaction point in the beam direction and within $1.0$~cm in the plane perpendicular to the beam. A PID algorithm, based on energy loss $dE/dx$ in the MDC, TOF information, and energy deposited in the EMC, is performed to identify electrons. An electron--positron pair is required for the selected events. In the decay $J/\psi \to e^+e^- \eta$, $\eta \rightarrow \pi^+\pi^-\pi^0$, the additional two charged tracks are assumed to be $\pi$ candidates without any PID requirement.

The photon candidates are reconstructed from the clusters of energy deposits in the EMC that are separated from the extrapolated positions of any charged tracks by more than $10$ degrees. The energy of each photon candidate  is required to be larger than $25$ MeV in the EMC barrel region ($|\cos\theta_{\gamma}| < 0.8$) or $50$ MeV in the EMC end-cap regions ($0.86<|\cos\theta_{\gamma}| < 0.92$), where $\theta_{\gamma}$ is the polar angle of the photon. To improve the reconstruction efficiency and energy resolution, the energy deposited in nearby TOF counter is taken into account. The photons reconstructed poorly  in the transition region between the barrel and the end-caps are discarded. The EMC timing is required to be within the range of $[0,700]$ ns  to suppress electronic noise and energy deposits unrelated to the event.

The selected charged tracks are constrained to originate from a common vertex point by requiring a successful vertex fit. In order to improve the resolution and further suppress the background,
a four-constraint (4C) kinematic fit that imposes overall momentum and energy conservation is implemented for the selected charged tracks and additional two photons under the hypothesis of $J/\psi\to e^+e^- (\pi^+\pi^-)\gamma\gamma$. The chi-square of the kinematic fit, $\chi^2_{\rm 4C}$, is required to be less than 100, which rejects about $30\%$ of the background events with a loss of the $10\%$ of the signal events.  If there are more than two good photons in an event, we try all the $\gamma\gamma$ combinations, and the one with the least $\chi^2_{\rm 4C}$ is chosen.
The kinematic variables after the 4C kinematic fit are used in the further analysis.
In the decay mode $\eta\to\pi^+\pi^-\pi^0$, the $\pi^0$ candidate is reconstructed with two selected photons by requiring $m_{\gamma\gamma}$ within the range of $[0.08, 0.16]$~GeV/$c^2$. The $\eta$ candidate is reconstructed with the selected $\gamma\gamma$ or $\pi^+\pi^-\pi^0$, respectively, and the corresponding masses ($m_{\gamma\gamma}$ and $m_{\pi^+\pi^-\pi^0}$) are required to be within the range $[0.45, 0.65]$~GeV/$c^2$.

\begin{figure}[!htp]
\begin{center}
\includegraphics[width=0.5\textwidth]{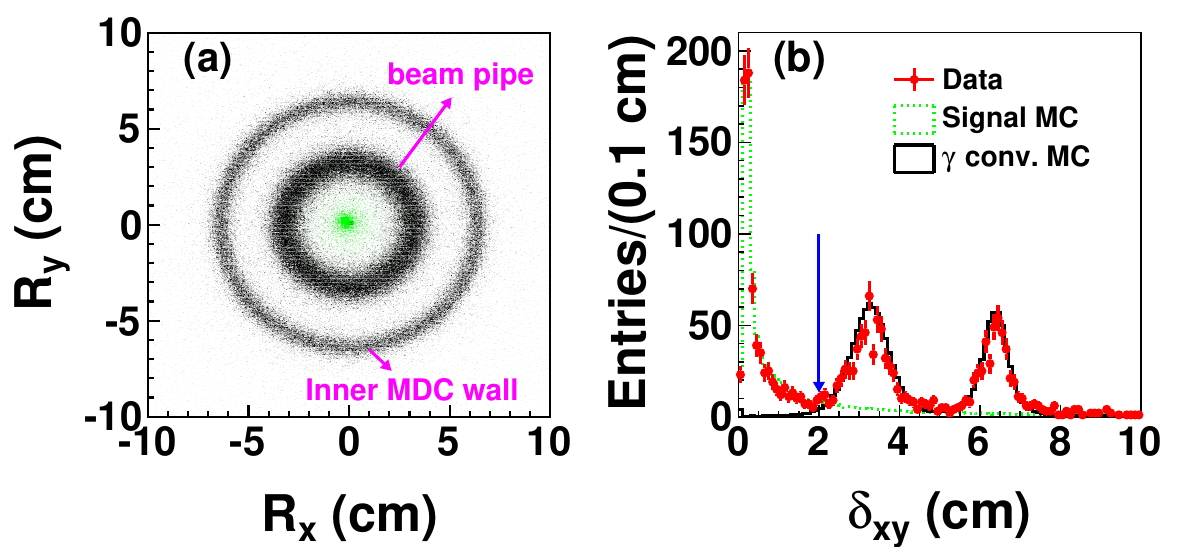}
\caption{(color online) (a) Scatter plot of $R_y$ versus $R_x$ for the simulated background MC events of $J/\psi \rightarrow \gamma \eta$ (black dot points) and signal MC events of $J/\psi\to e^+e^-\eta$ (green dot points), and (b) $\delta_{xy}$ distribution of signal MC (green dashed line), $\gamma$ conversion background MC events (black line) and data (red  dot error points). The requirement on $\delta_{xy}$ is shown by a solid blue arrow.}
\label{detaxy}
\end{center}
\end{figure}

With the above selection criteria, the peaking background, which contains an $\eta$ signal in the final state, is dominated by the events of the radiative decay $J/\psi \rightarrow \gamma \eta$ followed by the conversion of the radiative photon into an $e^+e^-$ pair in the detector material. In order to suppress this background, a photon-conversion finder algorithm~\cite{gamcon} is exploited to reconstruct the photon-conversion vertex point. The distance from the conversion vertex point to the origin in the $x-y$ plane, $\delta_{xy} = \sqrt{R_x^2+R_y^2}$, is used to separate the signal from the gamma conversion events, where $R_x$ and $R_y$ refer to the coordinates of the reconstructed vertex point along the $x$ and $y$ directions, respectively. The scatter plot of $R_y$ versus $R_x$ from the simulated $\gamma$ conversion background MC sample $J/\psi\to\gamma \eta$ and the signal MC sample $J/\psi\to e^+e^- \eta$ is shown in Fig.~\ref{detaxy}\,(a), where the circles with radius of $3.5$~cm and $6.5$~cm correspond to the positions of the beam pipe and inner wall of the MDC, respectively. The corresponding distributions of $\delta_{xy}$ from the signal and background MC samples, as well as data events are shown in Fig.~\ref{detaxy}\,(b).
The $\delta_{xy}$ is then required  to be less than $2$\,cm  to remove around $98\%$ of the $\gamma$ conversion events from $J/\psi \rightarrow \gamma \eta$ decay, while retaining about $80\%$ of the signal events $J/\psi\to e^+e^-\eta$.

In the decay mode $\eta \rightarrow \gamma \gamma$, the background is dominated by the non-peaking background from the QED processes $e^+e^-\to e^+e^- \gamma (\gamma)$ and $e^+e^- \rightarrow 3 \gamma$ in which one of the photons converts into an $e^+e^-$ pair. Since the $\eta$ meson decays isotropically, the cosine of the helicity angle ($\cos\theta_{\rm heli}$), defined as the angle between the direction of one of the photons and  $J/\psi$ direction in the $\eta$ rest frame, is expected to be uniformly distributed for signal events and to peak near $\cos\theta_{\rm heli} = \pm 1$ for the background from QED processes. Thus a requirement $|\cos\theta_{\rm heli}| < 0.9$ is implemented in the decay mode $\eta \rightarrow \gamma \gamma$ to suppress the non-peaking QED background.

\begin{figure}[hbtp]
\begin{center}
\includegraphics[width=0.5\textwidth]{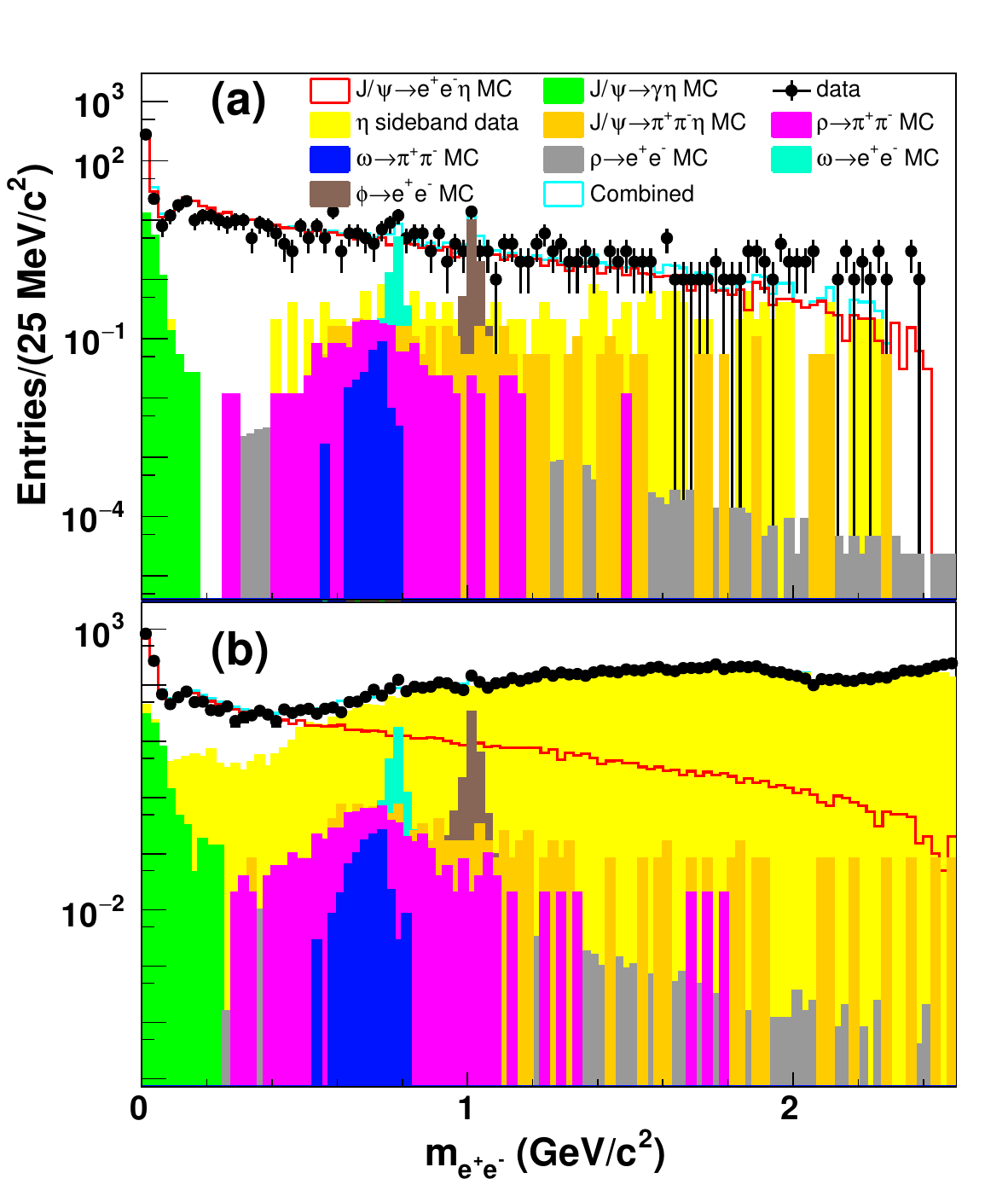}
\caption{(color online) Spectrum of $m_{e^+e^-}$ from data (black error dot points), signal MC (red), $\eta$ side-band data (yellow),  $J/\psi \to \gamma \eta$ MC (green),  $J/\psi \to \pi^+\pi^- \eta$ MC (orange),  $J/\psi \to \rho \eta$, $\rho \to \pi^+\pi^-$ MC  (pink), $J/\psi \to \rho \eta$, $\rho \to e^+e^-$ MC (brown),   $J/\psi \to \omega \eta$, $\omega \to \pi^+\pi^-$ MC  (blue), $J/\psi \to \omega \eta$, $\omega \to e^+e^-$ MC (teal),  $J/\psi \to \phi \eta$, $\phi \to e^+e^-$ MC (brown), and combined data of  MC and side-band (cyan) for the decay modes (a) $\eta \rightarrow \pi^+\pi^-\pi^0$ and (b) $\eta \rightarrow \gamma \gamma$.
}
\label{invmass}
\end{center}
\end{figure}

After applying the above selection criteria, the distribution of the di-electron invariant mass $m_{e^+e^-}$ of surviving events (within the $\eta$ signal region [0.51, 0.58]~GeV/$c^2$) is shown in Fig.~\ref{invmass}. Besides the EM Dalitz decay of interest, $J/\psi \rightarrow e^+e^-\eta$, small signals of $J/\psi \rightarrow V\eta$ ($V=\rho,\omega,\phi$) with $V$ subsequently decaying into $e^+e^-$ pair are observed.
Detailed MC studies indicate that the remaining peaking background is dominated by $J/\psi \rightarrow \gamma\eta$ with $\gamma$ converting into an $e^+e^-$ pair, which accumulates in the low region of the $m_{e^+e^-}$ distribution. There are also small contributions of the peaking background of $J/\psi \rightarrow \rho/\omega \eta$ with $\rho/\omega$ subsequently decaying into a $\pi^+\pi^-$ pair and the direct three body decay $J/\psi\to \pi^+\pi^- \eta$, in which the $\pi^+\pi^-$ are mis-identified as an $e^+e^-$ pair. The non-peaking background, which is smoothly distributed in the high mass region of the $m_{e^+e^-}$ distribution, 
is almost negligible in the decay mode $\eta\to\pi^+\pi^-\pi^0$, but sizable in the decay mode $\eta \rightarrow \gamma \gamma$ dominated by the radiative Bhabha $e^+e^-\rightarrow \gamma e^+e^-$ process. The distributions of signal and individual background components are also depicted in Fig.~\ref{invmass}.  Here, the peaking backgrounds are estimated with the MC simulation normalized according to the branching fraction quoted from the PDG~\cite{pdg}; the three body decay $J/\psi\to \pi^+\pi^- \eta$ is simulated in accordance with the amplitude of $J/\psi \rightarrow \pi^+\pi^-\eta'$~\cite{pwa}; the non-peaking backgrounds are estimated with the events of data in the $\eta$ sideband regions, which are defined as [0.42, 0.50]~GeV/$c^2$ and [0.59, 0.70]~GeV/$c^2$.

\subsection{Branching fraction measurement for the EM Dalitz decays \boldmath{$J/\psi \rightarrow e^+e^- \eta$}}
\label{BF}

In order to suppress the peaking background from $J/\psi\to V \eta$ with meson $V$ decaying into either the $e^+e^-$ or the $\pi^+\pi^-$ final state, the candidate events within regions of $0.65 < m_{e^+e^-} < 0.90$~GeV/$c^2$ or $0.96 < m_{e^+e^-} < 1.08$~GeV/$c^2$ are discarded. The number of remaining peaking background events, estimated by the MC simulation, for both $\eta$ decay modes after this requirement is summarized in Table~\ref{peakingback}. 

\begin{table}[htbp]
  \centering
  \caption{The remaining number of peaking background events in both the $\eta$ decay modes, where uncertainties are negligible.  }
  \renewcommand{\arraystretch}{2.0}
   \begin{tabular}{l | c |c}
  \hline \hline
 Decay process                                      & $\eta \to \gamma \gamma$       &  $\eta \to \pi^+\pi^-\pi^0$   \\ \hline  \hline
 $J/\psi \to \rho\eta$, $\rho \to \pi^+\pi^-$       &  $2.3$                 &   $0.6$                     \\
 $J/\psi \to \rho\eta$, $\rho \to e^+e^-$           &  $0.4$                 &   $0.1$                     \\
 $J/\psi \to \omega\eta$, $\omega \to \pi^+\pi^-$   &  $0.1$                 &   $0.0$                     \\
 $J/\psi \to \omega\eta$, $\omega \to e^+e^-$       &  $0.1$                 &   $0.0$                     \\
 $J/\psi \to \phi\eta$, $\phi \to e^+e^-$           &  $0.4$                 &   $0.1$                     \\
 $J/\psi \to \pi^+\pi^-\eta$                        &  $5.2$                 &   $1.9$                     \\
 $J/\psi \to \gamma \eta$                           &  $61.4$                &    $19.5$                     \\ 
\hline \hline
\end{tabular}
\label{peakingback}
\end{table}
  
In the decay mode $\eta \rightarrow \gamma \gamma$, a sizable non-peaking background, which is dominated by the radiative Bhabha events $e^+e^- \rightarrow \gamma e^+e^-$ and smoothly distributed in the high region of the $m_{e^+e^-}$ distribution, is suppressed by applying the further requirement  $p_{e^{\pm}} < 1.45$ GeV/c for  $m_{e^+e^-} > 0.5$~GeV/$c^2$, where $p_{e^{\pm}}$ is the momentum of the $e^{\pm}$ charged tracks. Other sources of peaking background are negligible in both $\eta$ decay modes.

\begin{figure}[hbtp]
\begin{center}
\includegraphics[width=0.5\textwidth]{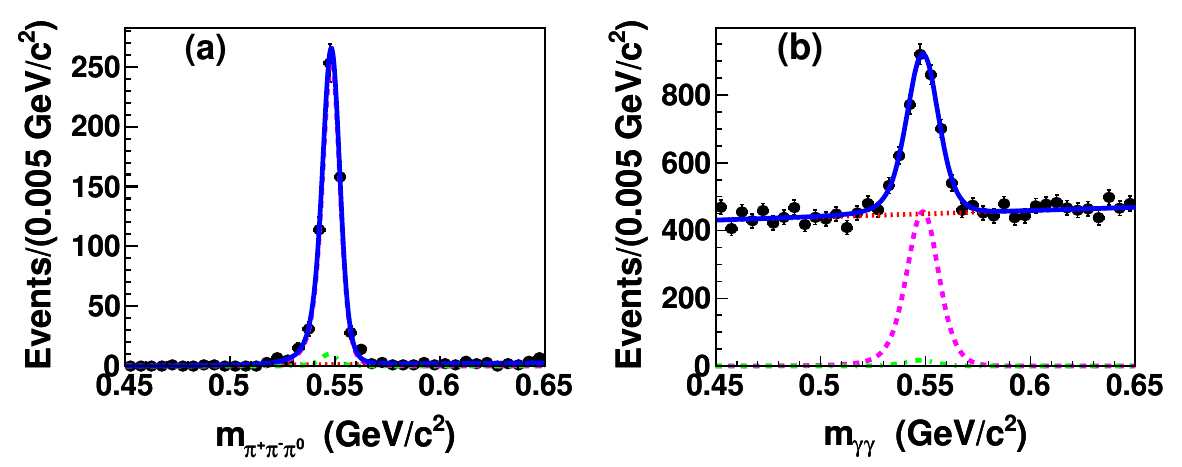}
\caption{(color online) Results of the unbinned ML fits to the distribution of (a) $m_{\pi^+\pi^-\pi^0}$ and (b) $m_{\gamma \gamma}$, respectively. The non-peaking background contribution is shown by a red dashed curve, the peaking background contribution by a green dashed curve, the signal distribution by a pink dashed curve and the total fit result by a solid blue curve.}
\label{projplot}
\end{center}
\end{figure}

To determine the signal yields, we perform an unbinned extended maximum likelihood (ML) fit to the  $m_{\gamma \gamma}$ and $m_{\pi^+\pi^-\pi^0}$ distributions, individually. In the fit, the probability density function (PDF) of the $\eta$ signal is described with the corresponding signal MC simulated shape convolved with a Gaussian function with parameters that are left free during the fit to take into account the resolution difference between the data and MC simulation.  The shape of the non-peaking background is described by a first order Chebyshev polynomial function with free parameters in the fit. The shape of the peaking background is described by that of MC simulation of the background $J/\psi \rightarrow \gamma \eta$, and the corresponding expected number of events is fixed during the fit. The ML fit yields $N_{\rm sig}=594.9 \pm 25.3$ and $1877.2 \pm 76.1$ events for the decay modes $\eta \rightarrow \pi^+\pi^-\pi^0$ and  $\eta \rightarrow \gamma \gamma$, respectively. The corresponding fit curves are shown in Fig.~\ref{projplot}. The statistical uncertainty of the extracted signal yield in $\eta \to \gamma \gamma$ slightly degrades compared to the previous BESIII measurement~\cite{xinkun}; this is because the ML fit to the $m_{\gamma \gamma}$ distribution is now performed in the full $m_{e^+e^-}$ range instead of $m_{e^+e^-} < 0.5$~GeV/$c^2$ range as required by the previous measurement to avoid the large contamination from the radiative Bhabha background.

\subsection{Transition form factor}
\label{tffmeasurement}
 Due to large contamination from the radiative Bhabha process in the high $m_{e^+e^-}$ region in the decay mode $\eta\to\gamma\gamma$, only the events from the $\eta \rightarrow \pi^+\pi^-\pi^0$ decay are used for the TFF study. The vicinities of $\omega$ and $\phi$ in the $m_{e^+e^-}$ distribution are also explored, and the resonant contribution of $J/\psi \to V \eta$, $V \to e^+e^-$ is considered as a signal in the TFF measurement.  Due to limited statistics in the high mass region as seen in Fig.~\ref{invmass}, the TFF is extracted bin-by-bin from the efficiency and branching fractions corrected signal yields for the bin sizes of $0.10$ GeV/$c^2$ between  $2m_e< m_{e^+e^-} < 1.10$ GeV/$c^2$, 
 $0.12$ GeV/$c^2$ between $1.10 < m_{e^+e^-} < 1.34$ GeV/$c^2$, $0.14$ GeV/$c^2$ between $1.34 < m_{e^+e^-} < 1.90$ GeV/$c^2$,  $0.16$ GeV/$c^2$ between $1.90 < m_{e^+e^-} < 2.06$ GeV/$c^2$ and $0.17$ GeV/$c^2$ in the remaining $m_{e^+e^-}$ regions with a total 20 bins, where $m_e$ is the mass of the electron. The signal yield in each bin of $m_{e^+e^-}$ is extracted by performing ML fits to the $m_{\pi^+\pi^-\pi^0}$ distribution as described in Sec.~\ref{BF}. The peaking background contribution from the $J/\psi \rightarrow \gamma \eta$ exists only in the first and second bins of $m_{e^+e^-}$. All the peaking background contributions including $J/\psi \rightarrow \gamma \eta$, $J/\psi \rightarrow \rho/\omega \eta$ ($\rho/\omega \rightarrow \pi^+\pi^-$) and $J/\psi \rightarrow \pi^+\pi^-\eta$ are estimated with the MC simulation and subtracted from the extracted signal yield from the ML fit in each bin.

\begin{figure}[hbtp]
\begin{center}
\includegraphics[width=0.5\textwidth]{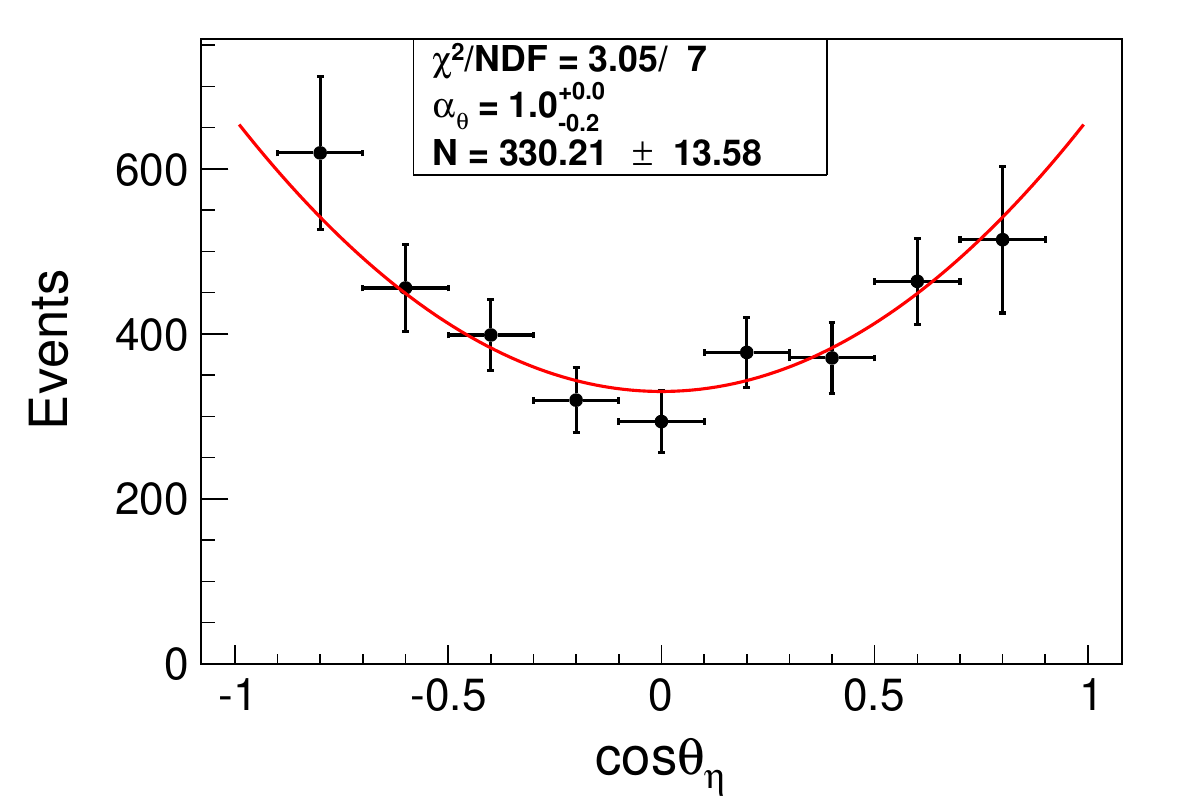}
\caption{Fit to the efficiency corrected signal yield versus $\cos\theta_{\eta}$ for data in the $\eta \to \pi^+\pi^-\pi^0$ decay mode.  The black dots with error bar are data, which include both statistical and systematic uncertainties, and the solid red curve shows the fit results. }
\label{nsigcostheta}
\end{center}
\end{figure}

 The signal efficiency for the TFF measurement is calculated by the signal MC sample generated according to the method discussed in Sec.~\ref{MCsimulation}, but with a constant TFF of $F_{J/\psi \eta}(m_{e^+e^-}^2)=1.0$. The  angular distribution parameter $\alpha_{\theta}$, used as an input parameter in this signal MC simulation, is evaluated after extracting the $\cos\theta_{\eta}$ dependent signal yield with a step size of 0.2 between  $-0.9 < \cos\theta_{\eta} < 0.9$ using a similar procedure of the ML fit mentioned above. Figure~\ref{nsigcostheta} shows the efficiency corrected signal yield versus $\cos\theta_{\eta}$ data and a fit with $\mathcal{N}(1+\alpha_{\theta}\cos^2\theta_{\eta})$, where $\mathcal{N}$ is a normalization constant and the efficiency for this study is evaluated after generating the simulated signal MC events with a flat distribution in $\cos\theta_{\eta}$. The angular distribution parameter $\alpha_{\theta}$ is determined to be $1.0^{+0.0}_{-0.2}$ with a condition of $0 \le |\alpha_{\theta}| \le 1.0$ to satisfy the theoretical constraints~\cite{alphaval}. 

\begin{figure}[hbtp]
\begin{center}
\includegraphics[width=0.5\textwidth]{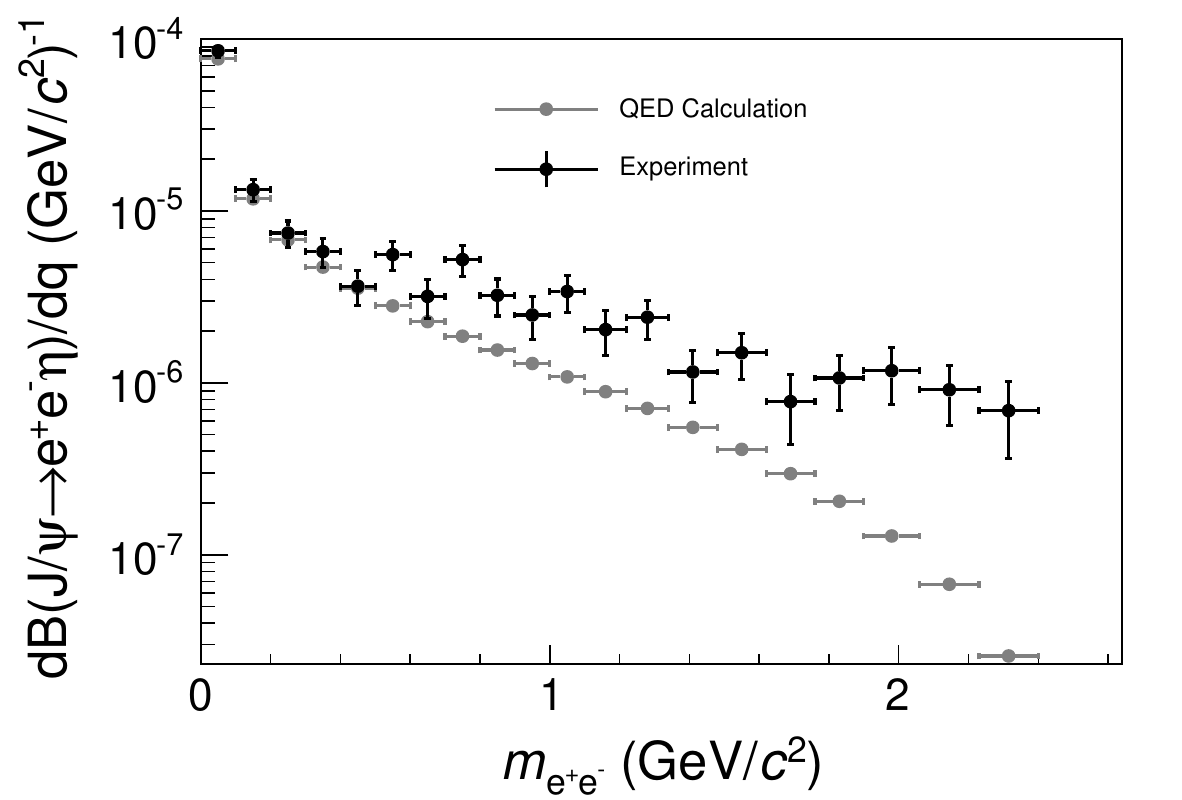}
\caption{Differential branching fraction $J/\psi \to e^+e^- \eta$  as a function of  $m_{e^+e^-}$. The black dots with error bars are experimental data, where the error bars include both statistical and systematic uncertainties, and the gray  dots with error bars are the MC prediction based on the pointlike QED calculation. }
\label{nsigQED}
\end{center}
\end{figure}

 Table~\ref{tfftab} summarizes the background subtracted fitted $N_{\rm sig}^i$ and branching fractions $\mathcal{B}(J/\psi \to e^+e^-\eta)^i$ for all 20 bins. The branching fraction of $J/\psi \to e^+e^- \eta$ is computed using
 \begin{equation}
\mathcal{B}(J/\psi \rightarrow e^+e^-\eta) = \frac{N_{\rm sig}}{N_{J/\psi}\cdot \mathscr{E} \cdot\mathcal{B}(\eta \rightarrow F)}
\label{branching_frac}
 \end{equation}
\noindent where $\mathscr{E}$ is the signal selection efficiency, $\mathcal{B}(\eta \rightarrow F)$ is the branching fraction of subsequent $\eta$ decays taken from the PDG~\cite{pdg} and $N_{J/\psi}=(1310.6 \pm 7.0)\times 10^{-6}$ is the number of $J/\psi$ events from Ref.~\cite{njps}. The distribution of $\mathcal{B}(J/\psi \to e^+e^-\eta)^i$ normalized to the $m_{e^+e^-}$ bin size superimposed with the QED predicted branching fractions, computed using the formula of Eq.~(\ref{eq:dgamman}), is shown in Fig.~\ref{nsigQED}.

\begin{table*}
  \centering
  \caption{Fitted $N_{\rm sig}^i$, differential branching fraction $\mathcal{B}(J/\psi \to e^+e^-\eta)^i$ and the TFF $|F(q^2)|^2$, described in Section~\ref{results},  for all 20 bins. The first uncertainty is statistical and the second systematic discussed in Section~\ref{syst}.}
  \renewcommand{\arraystretch}{2.0}
   \begin{tabular}{l c c c}
  \hline \hline
 $m_{e^+e^-}$~(GeV/$c^2$)  \hspace{0.2cm}  & $N_{\rm sig}^{i}$      \hspace{0.2cm}       & \hspace{0.2cm} $\mathcal{B}(J/\psi \to e^+e^-\eta)^i$ ($10^{-7}$) \hspace{0.2cm} & \hspace{0.2cm}  $|F(q^2)|^2$  \\ \hline  \hline
         [$2m_e$,   0.1]   &  $302.7 \pm 18.1  \pm 19.2$  &   $84.6 \pm 5.1 \pm 5.4$                            &  $1.11 \pm 0.07 \pm 0.07$ \\  
         {[0.1, 0.2]}       &  $60.9 \pm  7.8  \pm  3.9$  &   $13.3 \pm 1.7 \pm 0.8$                            &  $1.13 \pm 0.15 \pm 0.07$  \\
         {[0.2, 0.3]}       &  $40.4 \pm  6.6  \pm  2.6$  &   $7.4 \pm 1.2 \pm 0.5$                             &  $1.10 \pm 0.18 \pm 0.07$  \\
         {[0.3,  0.4]}      &  $32.0 \pm  5.7  \pm  2.0$  &   $5.8 \pm 1.0 \pm 0.4$                             &  $1.23 \pm 0.22 \pm 0.08$  \\
         {[0.4,  0.5]}      &  $20.6 \pm  4.6  \pm  1.3$  &   $3.7 \pm 0.8 \pm 0.2$                             &  $1.03 \pm 0.23 \pm 0.06$  \\
         {[0.5, 0.6]}       &  $31.6 \pm  5.7  \pm  2.0$  &   $5.6 \pm 1.0 \pm 0.4$                             &  $1.99 \pm 0.36 \pm 0.12$  \\
         {[0.6, 0.7]}       &  $18.2 \pm  4.5  \pm  1.3$  &   $3.2 \pm 0.8 \pm 0.2$                             &  $1.40 \pm 0.35 \pm 0.10$   \\ 
         {[0.7,  0.8]}      &  $29.8 \pm  5.7  \pm  1.9$  &   $5.2 \pm 1.0 \pm 0.3$                             &  $2.79 \pm 0.53 \pm 0.18$   \\ 
         {[0.8,  0.9]}      &  $19.1 \pm  4.5  \pm  1.2$  &   $3.2 \pm 0.8 \pm 0.2$                             &  $2.08 \pm 0.49 \pm 0.13$  \\ 
         {[0.9,  1.0]}      &  $14.4 \pm  3.9  \pm  0.9$  &   $2.5 \pm 0.7 \pm 0.2$                             &  $1.92 \pm 0.52 \pm 0.12$  \\ 
         {[1.0, 1.1]}       &  $19.8 \pm  4.6  \pm  1.2$  &   $3.4 \pm 0.8 \pm 0.2$                             &  $3.14 \pm 0.73 \pm 0.20$   \\
         {[1.1, 1.22]}      &  $14.6 \pm  4.2  \pm  1.0$  &   $2.5 \pm 0.7 \pm 0.2$                             &  $2.30 \pm 0.66 \pm 0.15$  \\ 
         {[1.22,  1.34]}    &  $16.8 \pm  4.1  \pm  1.1$  &   $2.9 \pm 0.7 \pm 0.2$                             &  $3.39 \pm 0.84 \pm 0.21$  \\ 
         {[1.34,  1.48]}    &  $9.7 \pm  3.2   \pm  0.6$  &   $1.6 \pm 0.5 \pm 0.1$                             &  $2.10 \pm 0.69 \pm 0.13$   \\
         {[1.48, 1.62]}     &  $12.4 \pm  3.6  \pm  0.8$  &   $2.1 \pm 0.6 \pm 0.1$                             &  $3.65 \pm 1.07 \pm 0.23$   \\
         {[1.62, 1.76]}     &  $6.3 \pm  2.7  \pm  0.6$   &   $1.1 \pm 0.5 \pm 0.1$                             &  $2.63 \pm 1.13 \pm 0.24$   \\
         {[1.76,  1.90]}    &  $9.1 \pm  3.1  \pm  0.6$   &   $1.5 \pm 0.5 \pm 0.1$                             &  $5.22 \pm 1.81 \pm 0.34$  \\ 
         {[1.90,  2.06]}    &  $10.2 \pm  3.7  \pm  0.6$  &   $1.9 \pm 0.7 \pm 0.1$                             &  $9.15 \pm 3.28 \pm 0.57$  \\  
         {[2.06, 2.23]}     &  $ 7.6 \pm   2.8 \pm  0.5$  &   $1.6 \pm 0.6 \pm 0.1$                             &  $13.54 \pm 5.05 \pm 0.87$   \\   
         {[2.23, 2.40]}     &  $ 5.7 \pm   2.7 \pm  0.4$  &  $1.2 \pm 0.6 \pm 0.1$                              &  $26.74 \pm 12.47 \pm 1.95$   \\  
\hline \hline
   \end{tabular}
\label{tfftab}
\end{table*}

 \subsection{The dark photon search in \boldmath{$J/\psi \rightarrow \gamma' \eta$} decays}
 The dark photon search is performed in the full $m_{e^+e^-}$ spectrum using the surviving event candidates within the $\eta$ mass window $[0.51, 0.57]$ GeV/$c^2$ of two $\eta$ decay modes.
 A series of unbinned extended ML fits to the $m_{e^+e^-}$ distribution is performed to determine the signal yields as a function of $m_{\gamma'}$ in the interval of $0.01\le m_{\gamma'} \le 2.40$ GeV/$c^2$.
 In the fit, the signal PDF is the sum of two Crystal Ball (CB) functions, which have common mean and width values, but opposite side tails. The parameters of the CB are extracted and extrapolated from the simulated signal MC events generated for  27 assumed $m_{\gamma'}$ points while assuming the width of the $\gamma'$ to be negligible in comparison to the experimental resolution.
 The background PDF is described by a composite function of polynomial and exponential functions, $f(m_{e^+e^-}) = c_0 \cdot m_{e^+e^-}+c_1 \cdot m_{e^+e^-}^2+e^{c_2 \cdot m_{e^+e^-}}$, for  $m_{\gamma'} < 0.2$ GeV/$c^2$, while a second order Chebyshev polynomial function is used in the remaining region.
 The signal selection efficiency and resolution vary in the range of $(5.0 - 37.0)\%$ ($(3.0-18.0)\%$) and $3-8$ MeV/$c^2$, for the decay mode of $\eta \rightarrow \gamma \gamma$ ($\pi^+\pi^-\pi^0$), respectively, depending on the momentum of the $e^{\pm}$ tracks.

\begin{figure}[htbp]
\begin{center}
\includegraphics[width=0.5\textwidth]{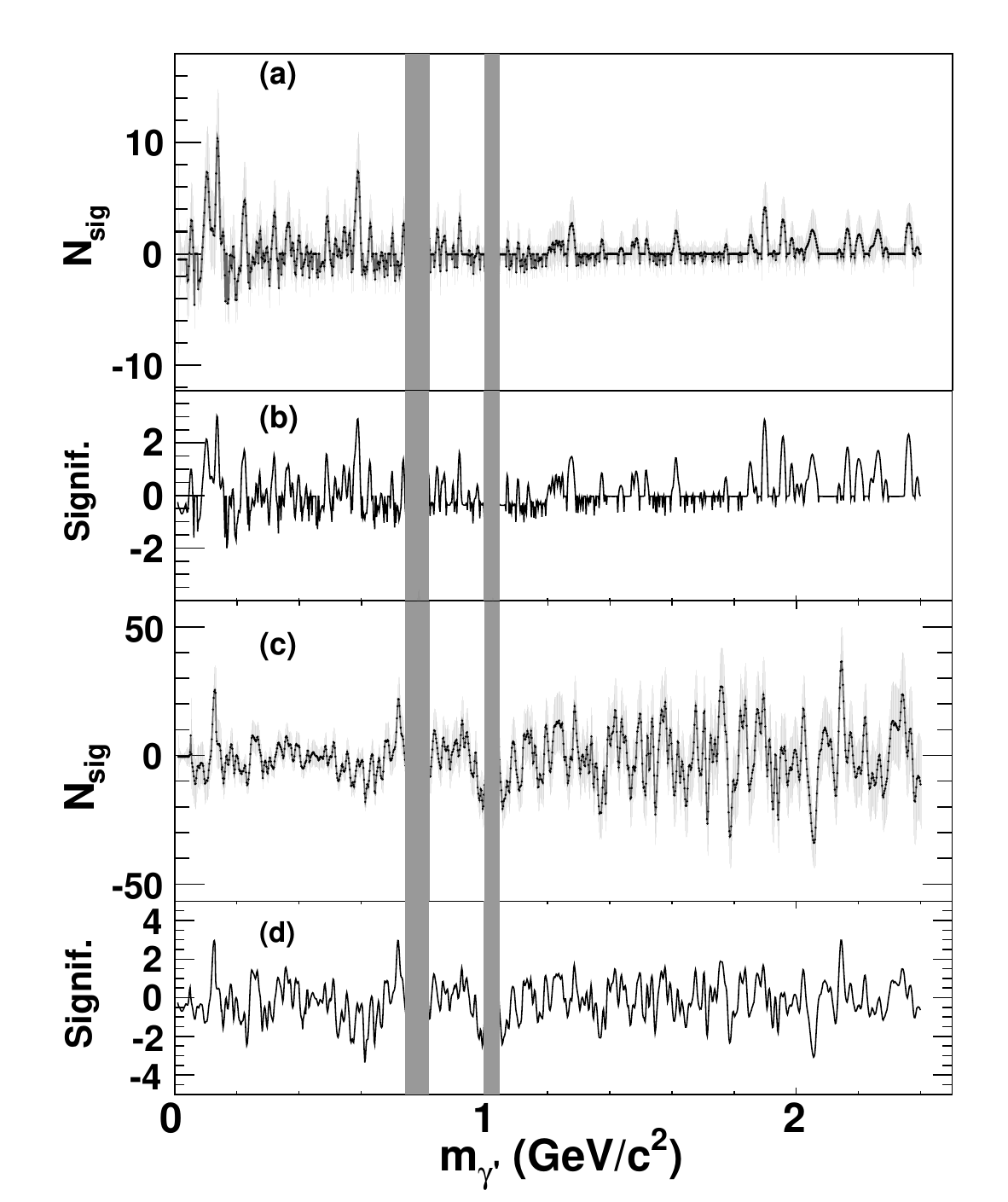}
\caption{Number of signal events and statistical signal significance  as a function of $m_{\gamma'}$ (a)-(b) for $\eta \rightarrow \pi^+\pi^-\pi^0$ decay and (c)-(d) for $\eta \rightarrow \gamma\gamma$ decay. The shaded regions of $\omega$ and $\phi$ resonances are excluded from the search. The asymmetric behavior in the high $m_{\gamma'}$ region of Fig. (a)-(b) is due to constraining the total PDF to be non-negative in the fit. }
\label{nsig}
\end{center}
\end{figure}

We search for the $\gamma'$ signal in steps of $2$ MeV/$c^2$ in the $m_{e^+e^-}$ distribution ranging from $10$ MeV/$c^2$ to $2.4$ GeV/$c^2$ excluding the vicinities of the $\omega$ and $\phi$ signals.
The parameters of the signal PDF are kept fixed, while the parameters of the background PDF, the number of signal events ($N_{\rm sig}$) and background events are determined by the fit.
In order to address the fit problem associated with low-statistics,  a lower bound of $N_{\rm sig}$ is imposed with a requirement that the total signal and background PDF remains non-negative~\cite{vindy}.  The statistical signal significance is computed  as $\mathcal{S} = {\rm sign}(N_{\rm sig})\sqrt{-2\ln(\mathcal{L}_0/\mathcal{L}_{max})}$, where $\mathcal{L}_{max}$  and $\mathcal{L}_0$ are the likelihood values when $N_{\rm sig}$ is left free and fixed at 0, respectively, and ${\rm sign}(N_{\rm sig})$ is the sign of $N_{\rm sig}$. The plots of $N_{\rm sig}$ and signal significance as a function of $m_{\gamma'}$ for both the $\eta$ decay modes are shown in Fig.~\ref{nsig}. The largest local significance is $2.92 \sigma$ at $m_{\gamma'} = 0.590$ GeV/$c^2$ in the $\eta \rightarrow \pi^+\pi^-\pi^0$ decay and $2.98 \sigma$ at $m_{\gamma'} = 2.144$ GeV/$c^2$ in the $\eta \rightarrow \gamma \gamma$ decay, which are less than $3\sigma$. Therefore, we conclude that no evidence of $\gamma'$ production is found in both the $\eta$ decay modes.

\section{Systematic uncertainty}
\label{syst}
Table~\ref{systfinal}  summarizes the sources of additive and multiplicative systematic uncertainties considered in this analysis, where the additive systematic uncertainties arise from the fit procedure including the signal and background modeling, as well as the bias of the fit procedure. The multiplicative systematic uncertainty  arises from the systematic uncertainty on the number of $J/\psi$ events, the branching fractions in the cascade decay and the event reconstruction and selection efficiencies.

\begin{table*}[hbtp]
  \centering
  \caption{Summary of systematic uncertainties. The systematic uncertainties  correlated between the decay modes $\eta\to\pi^+\pi^-\pi^0$ and $\eta\to \gamma\gamma$ are denoted by asterisks. Here \rm{\lq Negl.\rq} means negligible, and \rm{\lq ---\rq} means  the corresponding source of systematic uncertainty is not applicable in a particular decay process.  }
  \renewcommand{\arraystretch}{1.5}
  \begin{tabular}{l |c|c|c|c|c}
  \hline \hline
  Source    &  \multicolumn{3}{|c}{$J/\psi \to e^+e^- \eta$ }  & \multicolumn{2}{|c}{$J/\psi \to \gamma' \eta$} \\
\hline \hline
 & $\eta \rightarrow \gamma \gamma$ & $\eta \rightarrow 3 \pi$ & TFF measurement & $\eta \rightarrow \gamma \gamma$ & $\eta \rightarrow 3\pi$   \\
\hline \hline
\multicolumn{6}{c}{Additive systematic uncertainties (events) } \\
\hline
\hline
 Fixed PDFs                  & 8.50    & 0.9  & negligible    & 0.0 -- 1.0    & 0.0 -- 0.6    \\
 Non-peaking background     & 56.0   & 1.4  & 0.0 -- 0.6 & 0.0 -- 12.0     & 0.0 -- 5.0       \\
 Fit Bias                   & 2.0   & 0.1  & 0.1          & 0.1            & 0.1        \\

\hline
Total                          & 56.7  & 1.7  & 0.1-0.6     & 0.1 -- 12.0   & 0.1 -- 5.0     \\
\hline
\hline
\multicolumn{6}{c}{Multiplicative systematic uncertainties ($\%$)} \\
\hline
\hline
Charged tracks (* for $e$ track only)           & 2.4 & 4.4          & 4.4     & 2.4   & 4.4    \\
$e^{\pm}$  PID*                                 & 1.2 & 1.2          & 1.2     & 1.2   & 1.2   \\
Photon detection efficiency*                    & 2.0 & 2.0          & 2.0     & 2.0   & 2.0  \\
$\chi_{4C}^2$                                    & 0.9 & 0.9          & 0.9     & 0.9   & 0.9   \\
$\eta/\pi^0$ mass window requirement           & --- & 1.0          &  1.0     & 1.0   & 2.0 \\
Veto of $\gamma$ conversion*                       & 1.0 & 1.0          & $0.0-1.5$ & $0.0-1.5$  & $0.0-1.5$  \\
$\cos\theta_{\gamma}^{\rm hel}$                     & 1.9 & ---          & ---     & 1.9   & ---   \\
$e^{\pm}$ momentum                            & Negl. & ---        & ---     & ---    & ---   \\
TFF                                             & 1.0 &  1.5         & ---     & 1.0   & 1.5    \\
$\mathcal{B}(\eta \rightarrow \gamma \gamma)$   & 0.5 & ---          & ---     & 0.5   & ---   \\
$\mathcal{B}(\eta \rightarrow \pi^+\pi^-\pi^0)$ & --- & 1.2          & 1.2     & ---   & 1.2  \\
$\mathcal{B}(J/\psi \rightarrow \gamma \eta)$   & --- & ---          & 3.1     & 3.1   & 3.1 \\
$\mathcal{B}(\gamma' \rightarrow e^+e^-)$*      & --- & ---          & ---    & $0.0-14.0$ &  $0.0-14.0$   \\
Number of $J/\psi$  events*                          & 0.5 & 0.5          & 0.5     & 0.5   & 0.5   \\
 \hline
Total                                           & 4.2 & 5.6 &         $6.2 - 6.3$   & $5.2-15.0$ & $6.6-15.5$  \\

\hline \hline
\end{tabular}
\label{systfinal}
\end{table*}

In the measurements of the branching fraction of $J/\psi \rightarrow e^+e^- \eta$, the signal yields are determined by fitting the corresponding $m_{\gamma\gamma}$ and $m_{\pi^+\pi^-\pi^0}$ distributions, while in the TFF studies, the signal yields are extracted with the same fit procedure in the $m_{\pi^+\pi^-\pi^0}$ distribution only in different $m_{e^+e^-}$ bins.
The uncertainty associated with the signal model in the fit is studied by replacing the corresponding PDF to be the sum of two CB functions convolved with a Gaussian function, where the parameters of the CB functions are extracted from fits to the signal MC samples. The uncertainty associated with the peaking background is studied by varying its expected number of events within $\pm 1\sigma$ of the uncertainties in the fit, and observed to be negligible. The uncertainty associated with the non-peaking background is studied by replacing the corresponding PDF to be a second order Chebyshev polynomial function in the fit.  In the fit to search for the $\gamma'$ boson, the signal is modeled with the sum of two CB function whose parameters are extracted and extrapolated from the simulated MC samples at 27 $m_{\gamma'}$ points. The corresponding uncertainty is studied by changing the parameters of the CB functions within $\pm 1\sigma$ of their uncertainties, taking into account the correlation between the different parameters.
The uncertainty due to the background model is studied by changing the order of polynomial functions in the fit.
The changes in the signal yields due to the PDF parameters  are considered as the uncertainties.
To validate the reliability of fits, we produce a large number of pseudo-experiments, which are of the same statistics of data, and perform the same fit procedure in each pseudo-experiment.  The resultant average difference between the input and output signal yields is found to be very small and considered as one of the systematic uncertainties.

The tracking efficiency for charged pions is studied with the control sample of $J/\psi \rightarrow \pi^+\pi^-\pi^0$~\cite{tracking}. The difference between data and MC simulation is found to be $1\%$, and is considered as the systematic uncertainty of the charged pion. The efficiencies of tracking and PID for $e^{\pm}$ is explored with the control sample of radiative Bhabha events $e^+e^- \rightarrow \gamma e^+e^-$ in 2-dimensional bins of momentum versus polar angle. The resultant average differences on efficiency between data and MC simulation, 1.2\% for tracking  and 0.6\% for the PID, weighted according to the momentum and polar angle distribution of  the MC samples, are considered as the systematic uncertainties.

The photon reconstruction efficiency is studied with a control sample of $e^+e^- \rightarrow \gamma \mu^+\mu^-$, in which the momentum of the ISR photon is inferred from the four-momenta of the $\mu^+\mu^-$ pair~\cite{vindyphot}. The difference in the efficiency between data and MC simulation is smaller than 1\%, which is taken as the systematic uncertainty. In the decay mode $\eta\to\pi^+\pi^-\pi^0$, the uncertainty related with the $\pi^0$ mass window requirement is studied with a high statistics control sample of $J/\psi \rightarrow p\overline{p}\pi^0$ and is assigned to be 1\%. In the $\gamma'$ search, the uncertainty associated with the $\eta$ mass window requirement is studied with a control sample of $J/\psi \rightarrow p\overline{p}\eta$, and is assigned to be 1\%, too.

The uncertainty associated with the 4C kinematic fit is explored by utilizing a control sample of $J/\psi\to\pi^+\pi^-\pi^0$ in which the $\pi^0$ dominant decay modes of  $\pi^0\to\gamma\gamma$ and  $\pi^0\to\gamma e^+e^-$ are utilized to mimic the  $J/\psi\to e^+e^-\eta$ signal with subsequent decay modes $\eta\to\gamma\gamma$ and $\eta\to\pi^+\pi^-\pi^0$, respectively. The relative difference in efficiencies between data and MC simulation in the corresponding control samples is observed to be up to the level of $0.9\%$, and considered as the systematic uncertainty.

The control sample of $J/\psi \rightarrow \pi^+\pi^-\pi^0$, $\pi^0 \to \gamma e^+e^-$ is also utilized to evaluate the systematic uncertainty for the $\delta_{xy} <2$~cm requirement used to suppress the $\gamma$-conversion background. The simulated MC events for  $\pi^0\rightarrow \gamma e^+e^-$ are generated with a simple monopole approximation TFF, $F(m_{e^+e^-}^2)=1+a_{\pi} m_{e^+e^-}^2/m_{\pi^0}^2$, where  $m_{\pi^0}$ is the nominal $\pi^0$ mass and $a_{\pi} = 0.032 \pm 0.004$ is the slope parameter~\cite{pdg}. We extract the $\pi^0 \to \gamma e^+e^-$ signal from the data by performing a ML fit to the $m_{e^+e^-}$ distribution before and after the selection of $\delta_{xy} < 2$ cm requirement.  The corresponding differences in efficiencies, 1.0\% in the measurement of branching fraction of $J/\psi\to e^+e^-\eta$ and $(0.0-1.5)\%$ depending on $m_{e^+e^-}$ in the TFF measurement and $\gamma'$ search, are taken as the systematic uncertainties.

We similarly utilize the control sample $J/\psi \rightarrow \pi^+\pi^-\pi^0$, $\pi^0 \rightarrow \gamma \gamma$ to evaluate the systematic uncertainty due to the photon helicity angle requirement $|\cos\theta_{\rm heli}| < 0.9$ in the $\eta\to\gamma\gamma$ decay. The background in this control sample, $\pi^0 \to \gamma e^+e^-$,  has a flat shape in $m_{\gamma \gamma}$, and  is eliminated by performing a ML fit to the $m_{\gamma \gamma}$ distribution.  The uncertainty is evaluated to be up to the level of $1.9\%$ by comparing the efficiencies between the data and MC simulation, where the efficiency is the ratio of signal yields with and without this requirement applied. We extract the signal yield in $\eta \to \gamma \gamma$ decay by varying the requirement of $e^{\pm}$ momentum within one standard deviation of its statistical uncertainty, and one of the largest values of the relative difference between the signal yields is considered as the systematic uncertainty and found to be negligible.

In the branching fraction measurement of the $J/\psi\to e^+e^-\eta$ Dalitz decay, the signal MC samples used to evaluate the detection efficiency are generated  by following the TFF of  Eq.~(\ref{TFF}) with measured $\Lambda$ value of $2.56$ GeV/$c^2$ as described in Sec.~\ref{Data_sample}. Two alternative MC samples with values of the pole mass $\Lambda$ differing by $\pm 1\sigma$ are generated,  and the resulting largest relative difference in efficiencies, $1.5\%$ for the decay mode $\eta \rightarrow \pi^+\pi^-\pi^0$ and $1.0\%$ for the decay mode $\eta \rightarrow \gamma \gamma$, are considered as the systematic uncertainty.

The systematic uncertainties associated with the decay branching fractions of $\eta\to\pi^+\pi^-\pi^0$ and $\eta\to\gamma\gamma$ are taken from the PDG~\cite{pdg}. In the measurements of TFF and coupling strength between the dark sector and the SM $\epsilon$, the related uncertainty associated with the branching fraction of $J/\psi\to\gamma \eta$ is taken from the PDG~\cite{pdg}, and in the $\epsilon$ measurement, the uncertainty of the theoretical branching fraction $\mathcal{B}(\gamma' \rightarrow e^+e^-)$, dominated by the uncertainty of the R value~\cite{pdg}, varies in the range of $(0-14)\%$ depending upon the $m_{\gamma'}$~\cite{BAtoLL}. The uncertainty  of the number of $J/\psi$ events is determined to be $0.5\%$ using the inclusive hadronic events of the $J/\psi$ decays~\cite{njps}.

\section{Results}
\label{results}
 We compute the branching fraction of $J/\psi \rightarrow e^+e^-\eta$ in both decay modes of $\eta \rightarrow \gamma \gamma$ and $\eta \rightarrow \pi^+\pi^-\pi^0$ by using Eq.~(\ref{branching_frac}). The signal efficiency $\mathscr{E}$ is evaluated to be $26.2\%$ in decay mode $\eta \rightarrow \gamma \gamma$ and $13.8\%$ in decay mode $\eta \rightarrow \pi^+\pi^-\pi^0$ using the signal MC samples. The branching fraction of $J/\psi \to e^+e^- \eta$ is determined to be $(1.39 \pm 0.06 ({\rm stat}) \pm 0.07 ({\rm syst}))\times 10^{-5}$ in the decay mode $\eta \rightarrow \gamma \gamma$ and  $(1.45 \pm 0.06 ({\rm stat}) \pm 0.08 ({\rm syst}))\times 10^{-5}$ in the decay mode $\eta \rightarrow \pi^+\pi^-\pi^0$, where the first uncertainties are statistical and second systematic. A weighted average method~\cite{Cowan} is used to combine the two branching fraction measurements taking correlated~(shown by asterisks in Table~\ref{systfinal}) and uncorrelated 
systematic uncertainties into account. The combined $\mathcal{B}(J/\psi \to e^+e^- \eta)$ for both $\eta$ decay modes is $(1.42 \pm 0.04 ({\rm stat}) \pm 0.07 ({\rm syst}))\times 10^{-5}$. Compared with the previous measured value of $1.16 \pm 0.07(stat) \pm 0.06 (syst)$ from BESIII~\cite{xinkun}, the  total (statistical) uncertainty is reduced by a factor of $1.5$ ($1.8$). The central value of the measured branching fraction improves over the previous BESIII measurement~\cite{xinkun} due to taking into account of the $J/\psi$ polarization state in the $e^+e^-$ annihilation system during the signal MC simulation.  

\begin{figure}[hbtp]
\begin{center}
\includegraphics[width=0.5\textwidth]{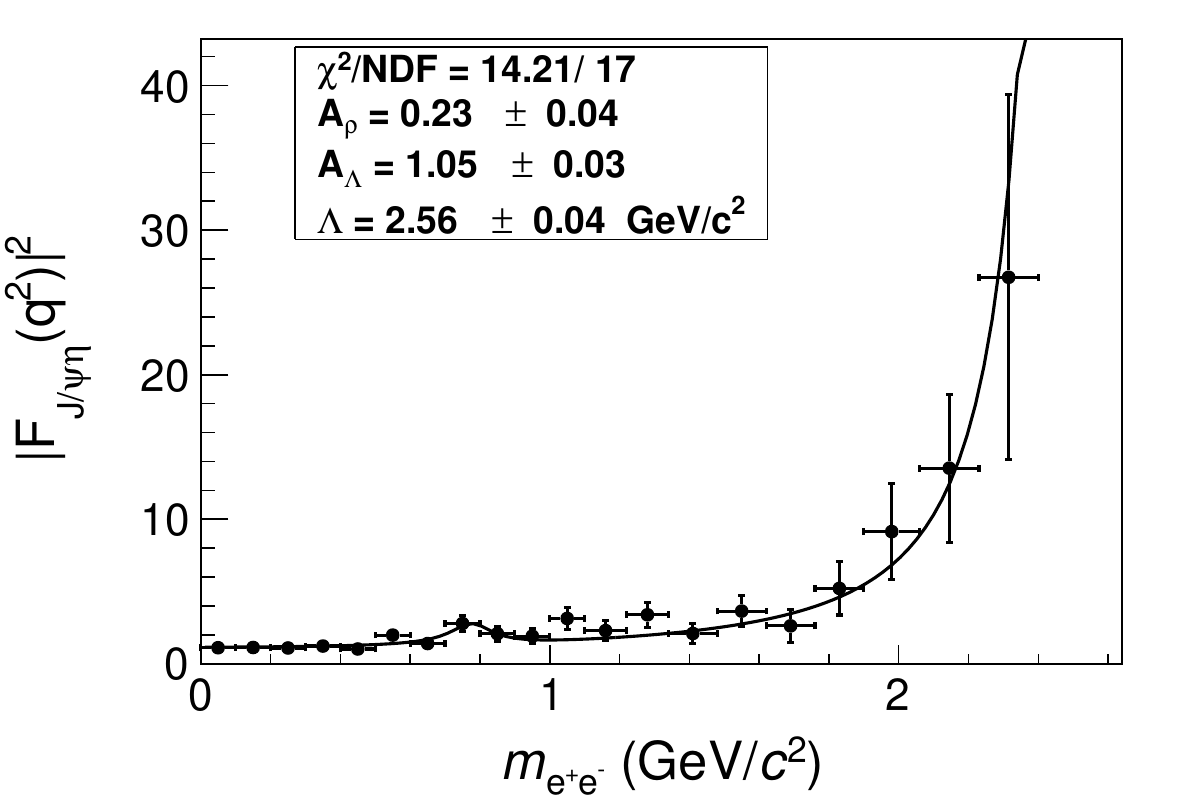}
\caption{Fit to the TFF versus $m_{e^+e^-}$ for data. The black dots with error bar are data, which include both statistical and systematic uncertainties, and the solid black curve shows the fit results.}
\label{tff}
\end{center}
\end{figure}

The TFF in each $m_{e^+e^-}$ bin is determined by dividing $\mathcal{B}(J/\psi \to e^+e^-\eta)^i$ by the integrated QED prediction in each $m_{e^+e^-}$ interval (Table~\ref{tfftab}). Figure~\ref{tff} shows a plot of the resultant TFF versus $m_{e^+e^-}$. A chi-square fit to the TFF versus $m_{e^+e^-}$ data is performed using  a modified multipole function of Eq.~(\ref{TFFcurve}), in which the contributions of the $\rho$ resonance and non-resonance are included, and the interference is neglected:
\begin{align}
 &  |F_{J/\psi \eta}(q^2)|^2 = |A_{\rho}|^2 \left(\frac{m_{\rho}^4}{(m_{\rho}^2-q^2)^2+\Gamma_{\rho}^2m_{\rho}^2}\right) \nonumber \\
 &\qquad {}  + |A_{\Lambda}|^2 \left(\frac{1}{1-q^2/\Lambda^2}\right)^2,
  \end{align}
\noindent where the mass and width of the $\rho$ resonance are fixed to the values in the PDG~\cite{pdg}. The statistical uncertainties and uncorrelated systematic uncertainty (between the different $m_{e^+e^-}$ bins) are considered when building the chi-square function. The fit curve is depicted in Fig.~\ref{tff}, too.  The statistical significance of the $\rho$ signal is  $4.3\sigma$ estimated with the change of chi-square values with and without $\rho$ signal included in the fit. We fit the TFF of data once again by including the interference between $\rho$ and non-resonant components. The resultant change on $\Lambda$, $0.023$ GeV/$c^2$, is taken to be one of the systematic uncertainties. We also fit the TFF of data without including the systematic uncertainty, the resultant change on $\Lambda$, $0.014$ GeV/$c^2$, is taken as another systematic uncertainty.
Finally, the pole mass is determined to be $\Lambda = 2.56 \pm 0.04({\rm stat}) \pm 0.03({\rm syst})$ GeV/$c^2$.

\begin{figure}[hbtp]
\begin{center}
\includegraphics[width=0.5\textwidth]{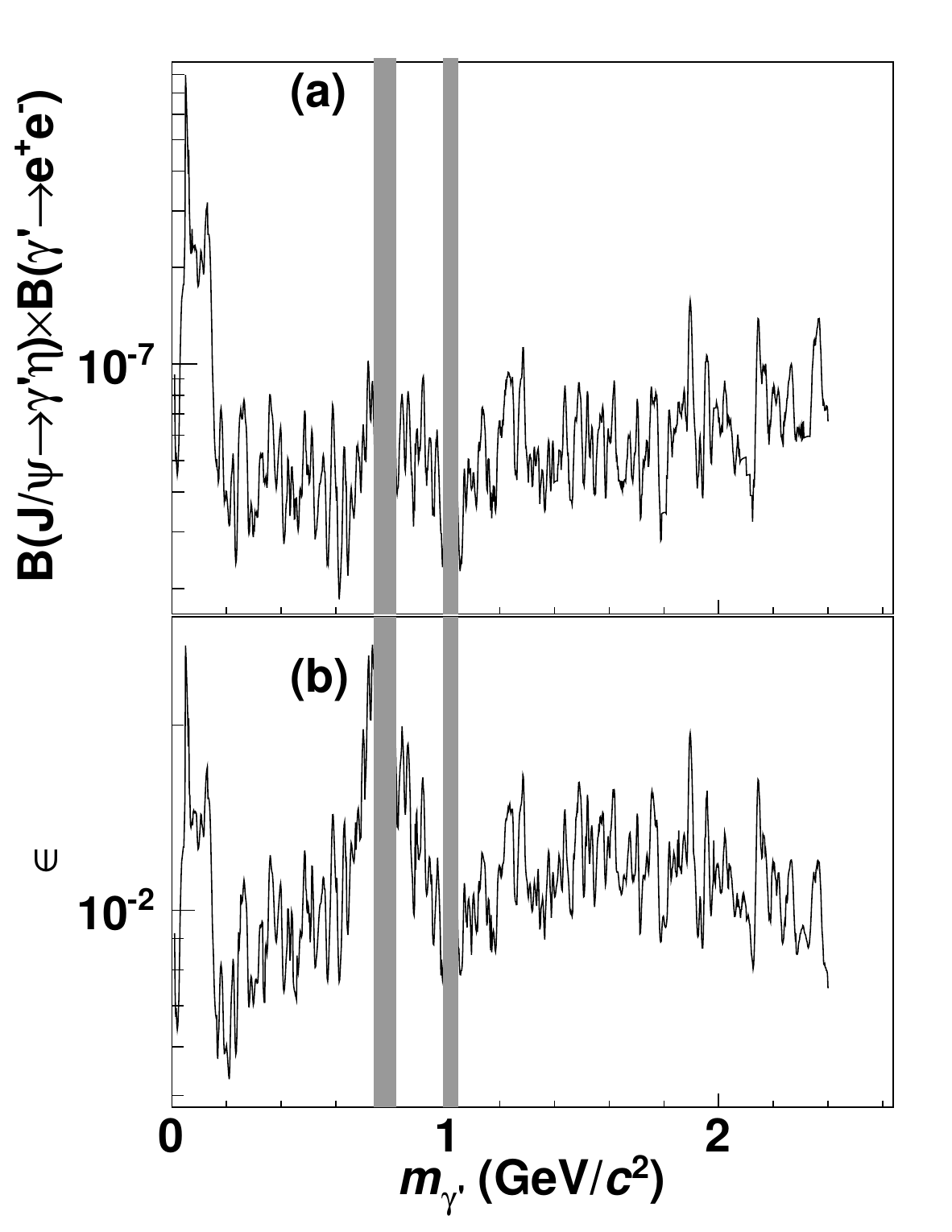}
\caption{ The combined upper limits at the 90\% C.L. on (a) product branching fraction $\mathcal{B}(J/\psi \rightarrow \gamma' \eta)\times \mathcal{B}(\gamma' \rightarrow e^+e^-)$ and (b) coupling strength ($\epsilon$) between  the SM and dark sector as a function of $m_{\gamma'}$ for both $\eta$ decay modes. The regions of $\omega$ and $\phi$ resonances shaded by gray lines are excluded from the $\gamma'$ search.
}
\label{bful}
\end{center}
\end{figure}

We compute the upper limits on the product branching fraction $\mathcal{B}(J/\psi \to \gamma' \eta)\times \mathcal{B}(\gamma' \to e^+e^-)$ at the $90\%$ confidence level (C.L.) as a function of $m_{\gamma'}$ using a Bayesian method after incorporating the systematic uncertainty by smearing the likelihood curve with a Gaussian function with a width of the systematic uncertainty. The combined result is obtained by adding the logarithm likelihoods of two $\eta$ decays by taking into account their correlated and uncorrelated systematic uncertainties. As shown in Fig.~\ref{bful}(a), the combined limits on product branching fraction  $\mathcal{B}(J/\psi \to \gamma' \eta)\times \mathcal{B}(\gamma' \to e^+e^-)$ vary in the range of $(1.9 - 91.1)\times 10^{-8}$ for $0.01 \le m_{\gamma'} \le 2.4$ GeV/$c^2$ depending on $m_{\gamma'}$ points.  The upper limit on $\mathcal{B}(J/\psi \to \gamma' \eta)$ at the 90\% C.L. at each $m_{\gamma'}$ point is computed by dividing the combined upper limit on the product  branching fraction  $\mathcal{B}(J/\psi \to \gamma' \eta)\times \mathcal{B}(\gamma' \to e^+e^-)$ by the expected dark photon decay branching fraction of $\gamma' \rightarrow e^+e^-$ obtained from Ref.~\cite{BAtoLL}.  We then compute the upper limits of the coupling strength between the dark sector and the SM $\epsilon$ at the 90\% C.L. as a function of $m_{\gamma'}$ using the Eq. (4.6) of Ref.~\cite{Reece}, where the TFF is given by Eq.~(\ref{TFF}) with $\Lambda = 2.56$ GeV/$c^2$. As shown in Fig.~\ref{bful}(b), the upper limits on $\epsilon$ at the 90\% C.L. vary in the range of $10^{-2} - 10^{-3}$ for $0.01 \le m_\gamma' \le 2.4$ GeV/$c^2$ depending on $m_{\gamma'}$.

\section{Summary}
In summary, with a data sample of $(1310.6 \pm 7.0)$ million $J/\psi$ events collected with the BESIII detector, we study the EM Dalitz decay of $J/\psi \to e^+e^-\eta$ and search for a dark photon in $J/\psi \to \gamma' \eta$ decay using two different $\eta$ decay modes $\eta\to\gamma\gamma$ and $\eta\to\pi^+\pi^-\pi^0$. The  branching fraction of $J/\psi \to e^+e^-\eta$ is measured to be $(1.42 \pm 0.04 ({\rm stat}) \pm 0.07 ({\rm syst}))\times 10^{-5}$, 
which supersedes the previous BESIII measurement~\cite{xinkun}. We present the first measurement of TFF as a function of $m_{e^+e^-}$ for the decay $J/\psi \to e^+e^-\eta$. The corresponding pole mass of the TFF  is determined to be $\Lambda = 2.56 \pm 0.04({\rm stat}) \pm 0.03({\rm syst})$ GeV/$c^2$  by fitting the  TFF versus $m_{e^+e^-}$ data with a modified TFF function.  No evidence of dark photon $\gamma'$ production is observed, and we set upper limits on the product branching fraction  $\mathcal{B}(J/\psi \to \gamma' \eta)\times \mathcal{B}(\gamma' \to e^+e^-)$
 at the $90\%$ C.L. to be in the range of $(1.9 - 91.1)\times 10^{-8}$ for $0.01 \le m_{\gamma'} \le 2.4$ GeV/$c^2$ depending on $m_{\gamma'}$. The upper limits on the coupling strength between the dark sector and the SM $\epsilon$ at the 90\% C.L. are also set at the level of $10^{-2} - 10^{-3}$, which are above the existing stringent experimental results~\cite{babar,NA48,bes}.

\section{Acknowledgement}
The BESIII collaboration thanks the staff of BEPCII, the IHEP computing center and the supercomputing center of USTC for their strong support. This work is supported in part by National Key Basic Research Program of China under Contract No. 2015CB856700; National Natural Science Foundation of China (NSFC) under Contracts Nos. 11235011, 11322544, 11335008, 11375170, 11275189, 11425524, 11475164, 11475169, 11625523, 11605196, 11605198, 11635010, 11705192; the Chinese Academy of Sciences (CAS) Large-Scale Scientific Facility Program; the CAS Center for Excellence in Particle Physics (CCEPP); Joint Large-Scale Scientific Facility Funds of the NSFC and CAS under Contracts Nos. U1332201, U1532257, U1532258, U1532102; CAS under Contracts Nos. KJCX2-YW-N29, KJCX2-YW-N45, QYZDJ-SSW-SLH003; 100 Talents Program of CAS; National 1000 Talents Program of China; INPAC and Shanghai Key Laboratory for Particle Physics and Cosmology; German Research Foundation DFG under Contracts Nos. Collaborative Research Center CRC 1044, FOR 2359; Istituto Nazionale di Fisica Nucleare, Italy; Joint Large-Scale Scientific Facility Funds of the NSFC and CAS; Koninklijke Nederlandse Akademie van Wetenschappen (KNAW) under Contract No. 530-4CDP03; Ministry of Development of Turkey under Contract No. DPT2006K-120470; National Natural Science Foundation of China (NSFC); National Science and Technology fund; The Swedish Research Council; U. S. Department of Energy under Contracts Nos. DE-FG02-05ER41374, DE-SC-0010118, DE-SC-0010504, DE-SC-0012069; University of Groningen (RuG) and the Helmholtzzentrum fuer Schwerionenforschung GmbH (GSI), Darmstadt; WCU Program of National Research Foundation of Korea under Contract No. R32-2008-000-10155-0.

\end{document}